\documentclass[aps,twocolumn,prx,superscriptaddress,footinbib]{revtex4-1}
\usepackage{amsmath,amssymb,bm}
\usepackage{graphicx}
\usepackage{epstopdf}
\usepackage{latexsym}
\usepackage[normalem]{ulem}
\usepackage[usenames,dvipsnames]{color}
\usepackage{hyperref}
\usepackage{natbib}
\begin{document}

\newcommand{\ns}{\mathcal{N}_{\mathrm{s}}}
\newcommand{\sinc}{\mathrm{sinc}}
\newcommand{\nb}{\mathcal{N}_{\mathrm{b}}}
\newcommand{\warn}[1]{{\color{red}\textbf{* #1 *}}}

\newcommand{\eqplan}[1]{{\color{blue}\textbf{equations:{ #1 }}}}

\newcommand{\figplan}[1]{{\color{blue}\textbf{figures: { #1 }}}}

\newcommand{\tableplan}[1]{{\color{blue}\textbf{tables: { #1 }}}}

\newcommand{\warntoedit}[1]{{\color{blue}\textbf{EDIT: #1 }}}

\newcommand{\warncite}[1]{{\color{green}\textbf{cite #1}}}

\newcommand{\Rev }[1]{{\color{blue}{#1}\normalcolor}} 
\newcommand{\Com}[1]{{\color{red}{#1}\normalcolor}} 
\newcommand{\RobCom}[1]{{\color{Mahogany}{#1}\normalcolor}} 
\newcommand{\JohnCom}[1]{{\color{purple}{#1}\normalcolor}} 
\newcommand{\MGCom}[1]{{\color{red}{MG: #1}\normalcolor}}

\newcommand{\mytitle}{ Verification of a many-ion simulator of the Dicke model through slow quenches across a phase transition}

%

\title{\mytitle}
\date{\today}

\author{ A. Safavi-Naini}
\thanks{These two authors contributed equally}
\affiliation{JILA, NIST and University of Colorado, 440 UCB, Boulder, CO 80309, USA}
\affiliation{Center for Theory of Quantum Matter, University of Colorado, Boulder, CO 80309, USA}
\author{ R. J. Lewis-Swan}
\thanks{These two authors contributed equally}
\affiliation{JILA, NIST and University of Colorado, 440 UCB, Boulder, CO 80309, USA}
\affiliation{Center for Theory of Quantum Matter, University of Colorado, Boulder, CO 80309, USA}
\author{J. G. Bohnet}
\affiliation{NIST, Boulder, CO 80305, USA}
\author{M. G\"{a}rttner}
\affiliation{JILA, NIST and University of Colorado, 440 UCB, Boulder, CO 80309, USA}
\affiliation{Center for Theory of Quantum Matter, University of Colorado, Boulder, CO 80309, USA}
\affiliation{Kirchhoff-Institut f\"{u}r Physik, Universit\"{a}t Heidelberg, Im Neuenheimer Feld 227, 69120 Heidelberg, Germany}
\author{K. A. Gilmore}
\affiliation{NIST, Boulder, CO 80305, USA}
\author{J. E. Jordan}
\affiliation{NIST, Boulder, CO 80305, USA}
\author{J. Cohn}
\affiliation{Department of Physics, Georgetown University, Washington, DC 20057, USA}
\author{J. K. Freericks}
\affiliation{Department of Physics, Georgetown University, Washington, DC 20057, USA}
\author{A. M. Rey}
\affiliation{JILA, NIST and University of Colorado, 440 UCB, Boulder, CO 80309, USA}
\affiliation{Center for Theory of Quantum Matter, University of Colorado, Boulder, CO 80309, USA}
\author{J. J. Bollinger}
\affiliation{NIST, Boulder, CO 80305, USA}

\begin{abstract}
We use a self-assembled two-dimensional Coulomb crystal of $\sim 70$ ions in the presence of an external transverse field to engineer a simulator of the Dicke Hamiltonian, 
an iconic model in quantum optics which  features a quantum phase transition between a superradiant/ferromagnetic and a normal/paramagnetic phase. We experimentally implement slow quenches across 
the quantum critical point and  benchmark the dynamics and the performance of the simulator through extensive theory-experiment comparisons which show excellent agreement. 
The implementation of the Dicke model in fully controllable trapped ion arrays can open a path for the generation of highly entangled
states useful for enhanced metrology and the observation of scrambling and quantum chaos in a many-body system.
\end{abstract}

\maketitle

\noindent{\it Introduction. }
Quantum many-body systems featuring controllable  coupled spin and bosonic degrees of freedom are becoming a powerful  platform for the realization of quantum simulators with easily tunable parameters.
These include for example cavity QED systems \cite{Leroux2010c,Hosten_2016,Ritsch2013,BGB10,BDR07,Baumann2010,Baumann2011,Klinder2015} and trapped-ion arrays \cite{Porras2004,Kim2009}. Most often,
these systems have been  operated in the far detuned regime  where the bosons do not play an  active role in the many-body dynamics and  instead  are used to mediate  spin-spin coupling between particles.
Great progress has been realized in this effective spin-model regime including the implementation of long range Ising models with and without an external transverse field and  the exploration of rich
physics  with them such as entanglement dynamics \cite{Leroux2010c,Hosten_2016,Blatt2012a,Jurcevic2014,Senko2015,Richerme2014,Islam2011,Bohnet2016,Garttner2016}, many-body localization \cite{Smith2015},
time crystals\cite{Zhanga2017} and  dynamical phase transitions ~\cite{Jurcevic2017,Zhang2017}. 
\begin{figure}[!]
 \includegraphics[width=0.5\textwidth]{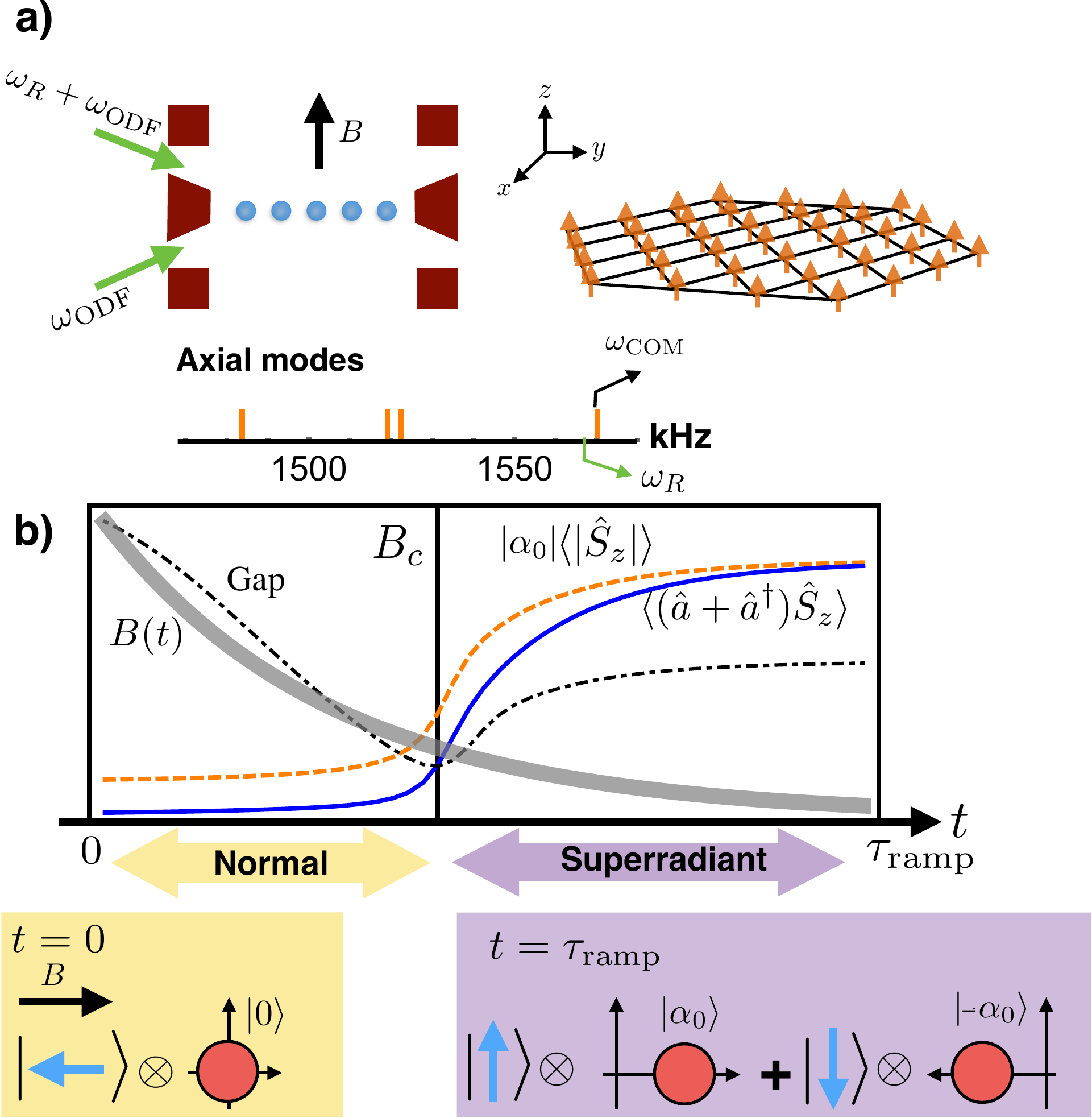}
\caption{Implementation and dynamical protocol. (a) The Dicke model is engineered with a Penning trap ion crystal of $N \sim 70$ ions  by applying an optical dipole force, resonant only with the center of mass
mode (which generates spin-phonon  interactions) and  resonant  microwaves (which generates the transverse field). The system is initially  prepared in  the normal phase where  all the spins point
along the transverse field and are decoupled from the  phonons. (b) As the  transverse field is   slowly turned off [using linear or  exponential ramp (shown here) profiles with  ramp time $\tau_{\rm ramp}$]
the system   enters  the superradiant phase after crossing the quantum critical point at  $B(t_{\rm crit})=B_c$ where the gap closes. The  superradiant phase with  macroscopic phonon  population, ferromagnetically
aligned spins  and  large spin-phonon entanglement is described by the order parameter $\langle(\hat{a}+\hat{a}^{\dagger})\hat{S}_z\rangle$, which is tracked closely by the re-scaled spin observable
$|\alpha_0|\langle|\hat{S}_z|\rangle$. (c) In the perfectly
adiabatic regime the ground state evolves from a separable spin-paramagnetic and vacuum photon Fock  state  into a macroscopic spin-phonon cat state: a superposition of two opposite  spin aligned and displaced-coherent
phonon states (with the sign of the superposition dictated by a parity symmetry, see SM).}
 \label{fig:schem}
\end{figure}
On the other hand, excluding  few  particle implementations  \cite{Pedernales2015,Lv2017,Lv2018rabi,Johnson2017,Kienzler2016,Monroe1996,Toyoda2015,Debnath2017,Raimond2001,Lamata2018}, the regime where  the bosonic degrees of
freedom  actively participate in the many-body  dynamics has remained largely unexplored.

In this work, we focus on this regime and report the implementation of a simulator of the Dicke model, an iconic model
in cavity QED which describes the coupling of a (large) spin and an oscillator, in a self-assembled  two-dimensional (2D) crystal of ions. The Dicke model is of broad
interest as it exhibits rich physics including quantum phase transitions and non-ergodic behaviour \cite{Altland2012}.
More recently it has gained renewed attention due to the implementation of the closely related Tavis-Cummings model in circuit QED \cite{Fink2009} and its 
realization in CQED experiments with ultracold  bosonic atoms  \cite{Baumann2010,Baumann2011,Klinder2015}. In the latter the Dicke model emerged as an effective Hamiltonian when one encodes a two-level system in  
two different momentum states of a Bose-Einstein condensate (BEC) coupled by the cavity field. Within this framework the normal to superradiant transition maps to a transition between a 
standard zero momentum BEC and a  quantum phase with macroscopic occupation of the higher-order momentum mode and the cavity mode.

While CQED experiments have used the intracavity light intensity and time of flight images to monitor the phase transition, here we 
instead probe the two distinct quantum phases of the Dicke model, by using  various controlled ramping protocols of a transverse field across the critical point (see Fig. \ref{fig:schem}). 
We benchmark the dynamics by experimentally measuring full distribution functions of the spin degrees of freedom and then comparing them with theoretical calculations. 
The spin observables also allow us to infer the development of spin-phonon correlations.

Our implementation of the Dicke model and corresponding observation of the phase transition in a trapped ion setup represents a complementary work with respect to the CQED platform 
and illustrates the power and universal nature of quantum simulation. It also opens a path for using the high level control and tunability  of trapped ions experiments for the  
generation of  highly entangled states suitable to quantum metrology in the near term future,  and for the  exploration of  regimes currently intractable to theory.


\noindent{\it Spin-Boson System. }
Our experimental system is comprised of a  2D single-plane array of laser-cooled
$^{9}$Be$^{+}$ ions in a Penning trap. The internal states forming the spin-1/2 system are the valence electron spin states in the Be$^+$ ion ground state which,
in the $4.46$~T magnetic field, are split by $124$~GHz \cite{Garttner2016,Bohnet2016,Sawyer2012,Biercuk2009}.
The interplay of the Coulomb repulsion  and  the  electromagnetic  confining  potentials supports a set of normal vibrational modes of the crystal \cite{Wang2013}, which we couple to the spin degrees of freedom
via a spin-dependent optical dipole force (ODF), generated by the interference of a pair of lasers   with beatnote frequency $\omega_R$ \cite{Sawyer2012}. The frequency $\omega_R$ is detuned from the
center-of-mass mode (COM) frequency,
$\omega_{\rm COM}$, by $\delta\equiv\omega_R-\omega_{\rm COM}$ (Fig. 1). The detuning is chosen to predominantly excite the COM mode  which  uniformly couples  all the ions in the crystal \cite{Bohnet2016}.
In the presence of an additional transverse field, generated by resonant microwaves, we implement  the Dicke Hamiltonian  \cite{Dicke1954,Garraway2011,Wall2017} :
\begin{eqnarray}
 \hat{H}^{\rm Dicke}/\hbar = -\frac{g_0}{\sqrt{N}} \left(\hat{a}+\hat{a}^{\dagger}\right)\hat{S}_z + B(t) \hat{S}_x - \delta \hat{a}^{\dagger}\hat{a}. \label{eq:HI}
\end{eqnarray}
in the frame rotating with $\omega_R$. The operator $\hat{a} (\hat a^\dagger)$ is the bosonic annihilation
(creation) operator for the COM mode, $B(t)$ is the time-varying strength of the applied transverse field, and $g_0$ represents the homogeneous coupling between each ion and the COM mode. Here, $\delta < 0$.
We have introduced the collective spin operators $\hat{S}_{\alpha} = (1/2)\sum_j \hat{\sigma}^{\alpha}_j$ where $\hat\sigma^{\alpha}_j$ is the corresponding Pauli matrix for $\alpha = x,y,z$ which acts on the $j$th ion.

The Dicke Hamiltonian exhibits a quantum phase-transition at $B_c=g_0^2/|\delta|$ in the thermodynamic limit, i.e. $N\to\infty$, \cite{Emary2003_PRL,Emary2003_PRE,Porras2013}, separating the normal ($B > B_c$ ) and superradiant ($B < B_c$ ) phases. 
The Hamiltonian remains unchanged under the simultaneous transformations $\hat S_z\to -\hat S_z$, $\hat{S}_y \to -\hat{S}_y$ and $\hat a \to -\hat a$. These are generated by the 
the parity operator $\hat{\Pi}=e^{i\pi(\hat{a}^{\dagger}\hat{a} + \hat{S}_x + \frac{N}{2})}$. 

In the strong-field regime of the normal phase, $B \gg B_c$, the spins and phonons decouple into a product state. When $|B|>|\delta|$  the corresponding ground state, $\vert \psi_{0,N/2}^{\rm{Nor}} \rangle$,
and low lying excitations, $\vert \psi_{n=1,2,\dots}^{\rm{Nor}} \rangle$, are
$\vert \psi_{n,N/2}^{\rm{Nor}} \rangle = \vert n\rangle \otimes \vert -N/2 \rangle_x$. We use $\vert n \rangle$ to denote Fock states  and
$\vert M\rangle_{\alpha=\{x,y,z\}}$ to denote the fully symmetric ($S=N/2$) eigenstates of $\hat{S}_{\alpha}\vert M\rangle_{\alpha}=M\vert M\rangle_{\alpha}$ with $-N/2\leq M\leq N/2$.

\begin{figure*}[!]
 \includegraphics[width=16cm]{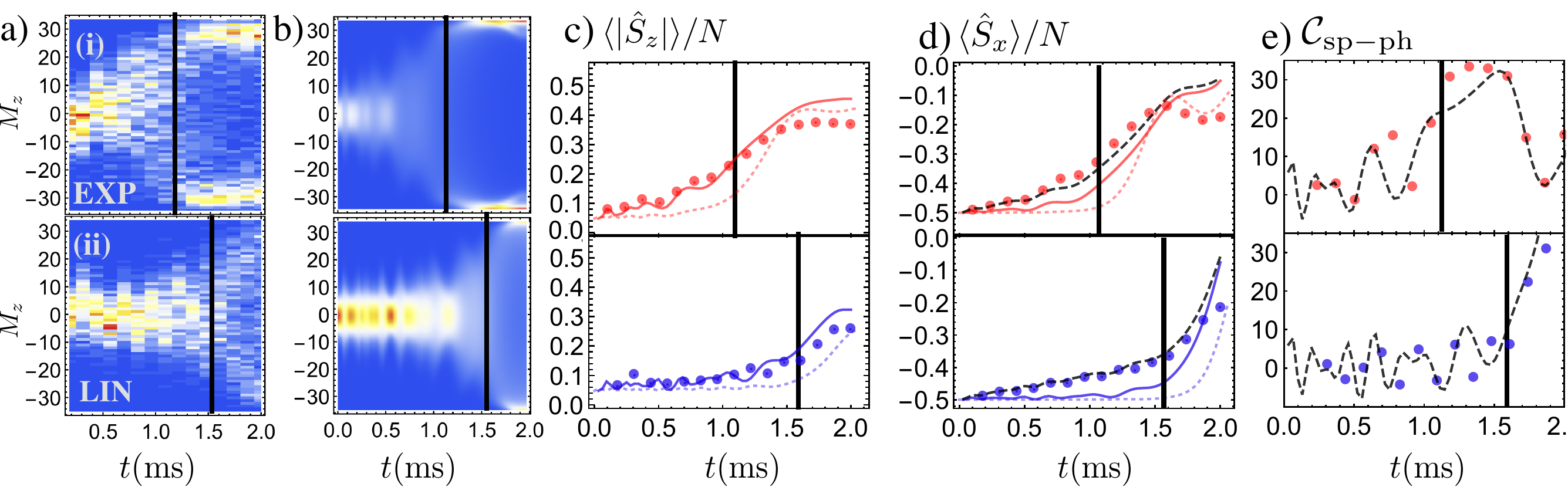}
 \caption{Benchmarking the simulator: Column (a) shows the experimentally measured  distribution function along $z$, and (b) the corresponding  theoretical simulations neglecting decoherence.
 Column (c) show the corresponding mean values of the magnetization $\langle |\hat{S}_z| \rangle$, (d) spin-projection $\langle \hat{S}_x \rangle$, and
 (e) $\mathcal C_{\rm sp-ph}\equiv \left \langle \left(\hat a + \hat a^\dagger\right) \hat S_y \right\rangle $.
 The filled circles are experimental measurements (statistical error is on the order of marker size), the colored solid and black dashed lines are the theory results without and with 
 dephasing [the latter curve is absent in panel (c) as the $z$-magnetization is less sensitive to this dominant source of decoherence] and the 
 colored dotted lines are the Lipkin model results. We indicate the time at which $B_c$ is reached in
 each ramp by a vertical line. The initial field is $B(t=0)/(2\pi) \approx 7.1$~kHz,  $g_0/(2\pi) \approx 1.32$~kHz,   $\delta/(2\pi) = -1$~kHz and $J/(2\pi)=1.75$~kHz.
 Respective ion numbers are $N = 68$ [EXP -- row (i)] and $N = 69$ [LIN -- row (ii)].} \label{fig:ExpResults_ExpecValues}
\end{figure*}

In the weak-field limit, $B \ll B_c$, of the superradiant phase, the spin and phonon degrees of freedom are entangled and the ground state is nearly degenerate in the thermodynamic limit. For a finite system it approaches 
$\vert \psi_{0, N/2}^{S} \rangle = \frac{1}{\sqrt{2}}\Big(\vert\alpha_0,0\rangle \otimes \vert N/2 \rangle_z \pm \vert-\alpha_0,0\rangle \otimes \vert -N/2 \rangle_z\Big)$ as $B \to 0$,
where we have introduced the displaced Fock states $\vert\alpha,n\rangle \equiv \hat{D}(\alpha)\vert n\rangle$ with $\hat{D}(\alpha) = e^{\alpha\hat{a}^{\dagger} - \alpha^*\hat{a}}$
the associated displacement operator \cite{Wunsche1991}. Here, the sign of the superposition is dictated by the parity symmetry: for even $N$ the ground-state will be the
symmetric superposition with $\langle e^{i\pi(\hat{a}^{\dagger}\hat{a} + \hat{S}_x + \frac{N}{2})} \rangle = 1$, while for odd $N$ the ground-state is the anti-symmetric superposition
with $\langle e^{i\pi(\hat{a}^{\dagger}\hat{a} + \hat{S}_x + \frac{N}{2})} \rangle = -1$.
In this weak-field regime the spins exhibit ferromagnetic order, 
characterized by the non-zero value of the order parameter $\vert \hat S_z \vert$,
while the phonon mode acquires a macroscopic occupation $\vert\alpha_0\vert^2$, where $\alpha_0 = g_0\sqrt{N}/(2\delta)$. 
The low-lying excitations correspond to  displaced Fock states, $\vert \psi_{n>0, N/2}^{S} \rangle$, if $\delta^2 < g_0^2$
and  to  spin-flips along $\hat z$, $\vert \psi_{0, M<N/2}^{S} \rangle $, if $\delta^2>g_0^2$.


{\it Slow quench dynamics. }
At the start of the experimental sequence  (see Fig.~\ref{fig:schem}) we prepare the initial spin state $\left|-N/2\right\rangle_{x}$ with the aid of a resonant  microwave pulse.
Doppler-limited cooling of the phonon degree of freedom leads to an initial phonon thermal state with mean occupation $\bar{n} \sim 6$.
For these parameters the system starts in the normal phase close to the ground-state.
 The transverse field is then quenched to zero (whilst the spin-phonon coupling and detuning are held constant)
according to two different profiles:
(i) Linear (LIN): $B(t)=B_0(1-t/\tau_{\mathrm{ramp}})$, and (ii) Exponential (EXP): $B(t) = B_0e^{-t/\tau}$. We set $\tau_{\mathrm{ramp}}=2$ms and $\tau \approx 600$~$\mu$s.

To characterize the performance of the simulator and the entrance into the superradiant phase, we  experimentally measure the full spin distribution along the
$z$ direction (Fig.~\ref{fig:ExpResults_ExpecValues}) by determining the global ion fluorescence scattered from the Doppler cooling laser on the cycling transition
for ions in $\left|\uparrow\right\rangle _{z}$ \cite{Martin2017_OTOC,Bohnet2016,Sawyer2012modeAndTempSpec,Biercuk2009}.  For repeated experimental trials we  infer the
state populations, $N_{\uparrow}$ and $N_{\downarrow}$   and calculate the spin-projection $M_z \equiv N_{\uparrow}-N/2$ for each experimental shot by counting
the total number of photons collected on a photomultiplier tube in a detection period, typically $5$~ms. Off-resonant light scattering from the ODF lasers is our
main source of decoherence  dominated by  single-particle dephasing at a rate  $\Gamma_{el}$ \cite{Uys2010}.


As noted above, the experimental implementation and corresponding numerical simulations were carried out with $N\approx70$ atoms. However, a well-defined cross-over between the normal and superradiant phases, 
signaled by a well-defined minimum in the energy gap between the ground and excited states of the same parity sector [see Fig. 1 (b)], appears for crystals larger than $N\gtrsim5$ (see SM).


Our theory-experiment comparisons are based on numerical solutions of the Dicke model dynamics combined with  thermal averaging. If decoherence is neglected  the spin degree of freedom is constrained to the $S=N/2$ manifold.
In this reduced Hilbert space we can exactly treat the quantum dynamics. 
Whilst for the non-negligible thermal phonon occupation in this experiment a classical treatment of the dynamics is sufficient to reproduce the 
measured observables, a complete formulation of the quantum dynamics becomes necessary for colder conditions, when thermal fluctuations are insufficient to drive dynamics and instead quantum correlations must be properly 
accounted for.
We observe good qualitative agreement  between the experimental spin probability distribution and the theoretically computed unitary dynamics as shown in
Figs. \ref{fig:ExpResults_ExpecValues}(a) and (b).  In particular, both show a
clear transition to a bimodal structure as the field strength is ramped down through $B_c$ (indicated by the black vertical line in each plot), with some ``smearing" due to the thermal occupation of the phonons.

To quantitatively determine the performance of the simulator, we plot the evolution of the effective order parameter $\langle \vert \hat{S}_z \vert \rangle/N$ (experimental values are 
extracted from the measured distribution) in
Fig.~\ref{fig:ExpResults_ExpecValues}~(c), which clearly builds up as one crosses $B_c$. The transition is not  abrupt and instead exhibits small amplitude oscillations, most clearly evident in the theoretical calculations, 
which reflect the active role of the phonons
given our initial finite thermal phonon occupation. In particular, our numerical simulations show a dependence of the oscillation amplitude 
on the initial phonon occupation (see SM). However, the frequency
of the phonon oscillations is difficult to determine and interpret, as it depends on the complex interplay between the magnitude of the initial phonon occupation and the changing transverse field.
We contrast this behavior with the case when the phonons can be adiabatically eliminated and  realize an effective spin Lipkin model,
$\hat{H}^{\rm LM}/\hbar = (J/N)\hat{S}_z^2 + B(t)\hat{S}_x$ where $J = g_0^2/\delta$. The Lipkin model dynamics features a sharper increase in magnetization after the critical point, and significant disagreement with the experimental observations.

To further benchmark the simulator we carry out similar measurements of the spin distribution along the $x$ direction, extracted  by applying a global $\pi/2$ pulse before the fluorescence measurement.
Fig.~\ref{fig:ExpResults_ExpecValues}~(d) shows the  mean-value of the  spin-projection $\langle\hat{S}_x\rangle$.
We observe $x$-depolarization as the system exits the normal phase. The Lipkin model dynamics also  exhibits a sharper depolarization  across $B_c$ than the one seen in the experiment.  In this case, however,
we do observe  deviations between the experiment and the ideal theory. The reason is that unlike the $z$-magnetization, this observable is strongly
affected by dephasing. Since treating the full spin-boson system in the presence of decoherence is computationally challenging, we model the effect of
dephasing as $\langle \hat{S}_x\rangle \rightarrow \langle \hat{S}_x \rangle e^{-\Gamma t}$ and $\langle \hat{S}_z\rangle\to \langle \hat{S}_z \rangle$ where $\Gamma=\Gamma_{el}/2$,
which is asymptotically valid in the $B \gg B_c$ and  $B \ll B_c$ limits \cite{Huelga1997}. We can determine $\Gamma_{el}$ experimentally when $B=0$, and we find $\Gamma_{el} \approx 120$~s$^{-1}$.
However, at large $B$, most clearly evidenced in the LIN protocol, the demagnetization
is faster than this estimate, and is consistent with $\Gamma_{el} = 280$~s$^{-1}$\footnote{This dephasing could be a result of the experimental system going beyond the Lamb-Dicke regime,
which is implicitly assumed in the derivation of the Dicke Hamiltonian Eq.~(1).}.
For both ramps  we observe excellent agreement to the experiment when  dephasing is accounted for.


Although measuring the phonon population might be possible  following the protocol reported in Ref.~\cite{Gilmore2017},  we instead infer the build-up of spin-phonon correlations  from the time evolution of the
spin observable $\langle \hat{S}_x \rangle$. Specifically, we assume the dynamics of the system are captured by the Lindblad master equation for the density matrix of the spin-phonon system $\hat{\rho}$,
\begin{equation}
 \frac{d\hat{\rho}}{dt} = -\frac{i}{\hbar} \left[ \hat{H}^{\mathrm{Dicke}}, \hat{\rho} \right] +\frac{\Gamma_{el}}{2}\sum_{i=1}^N \left( \hat{\sigma}^z_i \hat{\rho} \hat{\sigma}^z_i - \hat{\rho} \right) ,
\end{equation}
where single-particle dephasing is taken to be the dominant decoherence mechanism. From the master equation we derive the equation of motion $\frac {d}{dt}\langle \hat{S}_x \rangle$, and rearrange to obtain the 
relation (see SM)
\begin{equation}
 \mathcal C_{\rm sp-ph}\equiv  \langle \left(\hat a + \hat a^\dagger\right) \hat S_y \rangle \equiv \frac{\sqrt{N}}{g_0} \left( \Gamma_{el} \langle \hat S_x \rangle + \frac {d}{dt}\langle \hat{S}_x \rangle \right) .
\end{equation}
We extract the spin-phonon correlation from the experimental data by evaluating the RHS of the above expression, and calculating the time-derivative numerically with a one-sided derivative. The results are plotted 
in Fig.~\ref{fig:ExpResults_ExpecValues}(e). We use  the same value of $\Gamma_{el}$ as in Fig.~\ref{fig:ExpResults_ExpecValues}(d). The results are compared
with a theoretical calculation of $C_{\mathrm{sp-ph}}$ [again modelling dephasing using $\langle \hat{S}_x \rangle_{\Gamma} \equiv \langle \hat{S}_x \rangle_{\Gamma=0}e^{-\Gamma t}$]. 
In principle, the correlator vanishes when evaluated for the ground-state at any field strength. However, for these slow quenches it acquires a finite value, which in particular grows in the superradiant phase, 
due to population of excited states. This is attributable due to diabatic excitations created during the ramping protocol or the initial thermal phonon ensemble. Thus, while the correlation $\mathcal C_{\rm sp-ph}$ 
shows similar dynamical features observed in the other observables, it gives an alternative insight into the excitations created during the ramp.

While we have used the two ramp profiles to benchmark the experiment, we note that the EXP ramp has more utility in preparing a final state close to the expected ground-state $\vert \psi_{0,N/2}^{S} \rangle$ 
in the superradiant phase. For instance, the EXP ramp produces a clearer bimodal structure in the spin probability distribution along $z$, and associated larger mean absolute spin projection 
$\langle \vert \hat{S}_z \vert \rangle$. Future experiments could improve assesment of the adiabaticity of the quench protocols by measuring 
any coherences present between the different spin components, as discussed below.

Accounting for spin-phonon entanglement will be key to properly diagnose the  generated many-body quantum state. 
For example, tracing out the phonons from $\vert \psi_{0,N/2}^{S} \rangle$ will exponentially suppress the coherence between the spin states $\vert \pm N/2 \rangle_z$ 
(see SM).  To benchmark the performance of the adiabatic dynamics it is  then highly desirable to  first perform a  protocol to disentangle the spins and phonons and only after it 
characterize the state by  independently measuring the spins and the phonons without information loss.


To disentangle spin and phonons we propose to instantaneously quench the detuning
$\delta \rightarrow \delta^{\prime} = 2\delta$ at the end of the LAA ramp ($B \rightarrow 0$) and then let the system evolve for a time $t_d=\pi/\delta^\prime$. At $t_d$  the phonons are coherently
displaced by $ - g_0\sqrt{N}/(2|\delta|) \langle S_z\rangle$ back to the origin, while the spins only acquire an irrelevant global phase \cite{Wall2017}.
The resulting disentangled state ideally becomes 
$(1/\sqrt{2})[\vert+\alpha_0,0\rangle\vert+ N/2\rangle_z + \vert-\alpha_0,0\rangle\vert- N/2\rangle_z] \rightarrow (1/\sqrt{2})\vert0\rangle\otimes[\vert+ N/2\rangle_z + \vert- N/2\rangle_z]$  which has maximal spin coherence.

{\it Summary and discussion.} We have reported the experimental realization of a simulator of the Dicke model with a 2D ion crystal of $\sim 70$ ions and verified its dynamics through extensive theory-experiment
comparisons. Our trapped-ion simulator provides a complementary approach to related realizations in cold atoms \cite{Baumann2010,Baumann2011,Klinder2015}, 
which is a key step in benchmarking quantum simulators which go beyond the capacity of classical computation.
Our realization of a many-ion simulator of the Dicke model also paves the way for future investigation of dynamical phase transitions \cite{Jurcevic2017,Zhang2017},
quantum chaos and fast scrambling via out-of-time order correlation measurements \cite{Shenker2014,Kitaev2015,Maldacena2016,swingle2016measuring,Garttner2016}. Moreover, the tunability of the trapped-ion setup opens
the possibility of investigating more general spin-boson models \cite{Athreya2017sync}, in particular by operating beyond the uniform coupling regime or the preparation of states outside the fully symmetric Dicke manifold.

The slow quench protocols demonstrated above present a path to generate highly entangled states useful for quantum enhanced metrology \cite{Feldmann2018}. Cat-states are a useful metrological resource as they are 
composed of a coherent superposition of states that are macroscopically displaced in phase-space, leading to quantum-enhanced phase-sensitivity up to the Heisenberg limit \cite{Bollinger1996,Toscano2006}. 
In particular, the spin-boson cat-state $\vert \psi_{0,N/2}^{S} \rangle$ would be a metrological resource for sensing collective spin rotations \cite{Bollinger1996}, 
motional rotation \cite{Campbell2017,Johnson2017}, and coherent displacements for force sensing applications \cite{Penasa2016}.
This could be achieved by using smaller systems (e.g., $N \sim 20$), reducing the initial thermal population of the phonon mode, and shifting the detuning $\delta$ away from $B_c$, which increases the minimum energy gap at the critical point, and consequently the characteristic time-scale to remain adiabatic (see SM).
We expect this regime will be accessible in the near term future in part due to the successful implementation of
electromagnetic induced transparency cooling \cite{futurePaper}.

%


\begin{acknowledgments}
The authors acknowledge fruitful discussions with J. Marino, M. Holland and K. Lehnert.
A.~M.~R acknowledges support from Defense Advanced Research Projects Agency (DARPA) and Army Research Office grant W911NF-16-1-0576, NSF grant PHY1521080, JILA-NSF grant PFC-173400, and the Air Force Office of Scientific
Research and its Multidisciplinary University Research Initiative grant FA9550-13-1-0086. M.G. acknowledges support from the DFG Collaborative Research Center SFB1225 (ISOQUANT). J.~E.~J. also acknowledges support from
Leopoldina Fellowship Programme. JKF and JC acknowledge support from NSF grant PHYS-1620555. In addition, JKF acknowledges support from the McDevitt bequest at Georgetown. Financial support from NIST is also acknowledged.
\end{acknowledgments}

\bibliography{library}

\clearpage

\onecolumngrid
\vspace{\columnsep}
\begin{center}
\textbf{\large Supplemental Material: Verification of a many-ion simulator of the Dicke model through slow quenches across a phase transition}
\end{center}
\vspace{\columnsep}
\twocolumngrid

\setcounter{equation}{0}
\setcounter{figure}{0}
\setcounter{table}{0}
\setcounter{page}{1}
\makeatletter
\renewcommand{\theequation}{S\arabic{equation}}
\renewcommand{\thefigure}{S\arabic{figure}}
\renewcommand{\bibnumfmt}[1]{[S#1]}
\renewcommand{\citenumfont}[1]{S#1}

\section{Finite size effects in observing the phase transition}
The quantum phase transition of the Dicke model only truly emerges in the thermodynamic limit $N\to\infty$ \cite{Emary2003_PRL,Emary2003_PRE}. It is thus important to consider the relevance 
of finite size effects, specifically pertaining to the number of ions $N$ and thus the collective spin length $S=N/2$. 

\begin{figure*}[!]
 \includegraphics[width=16cm]{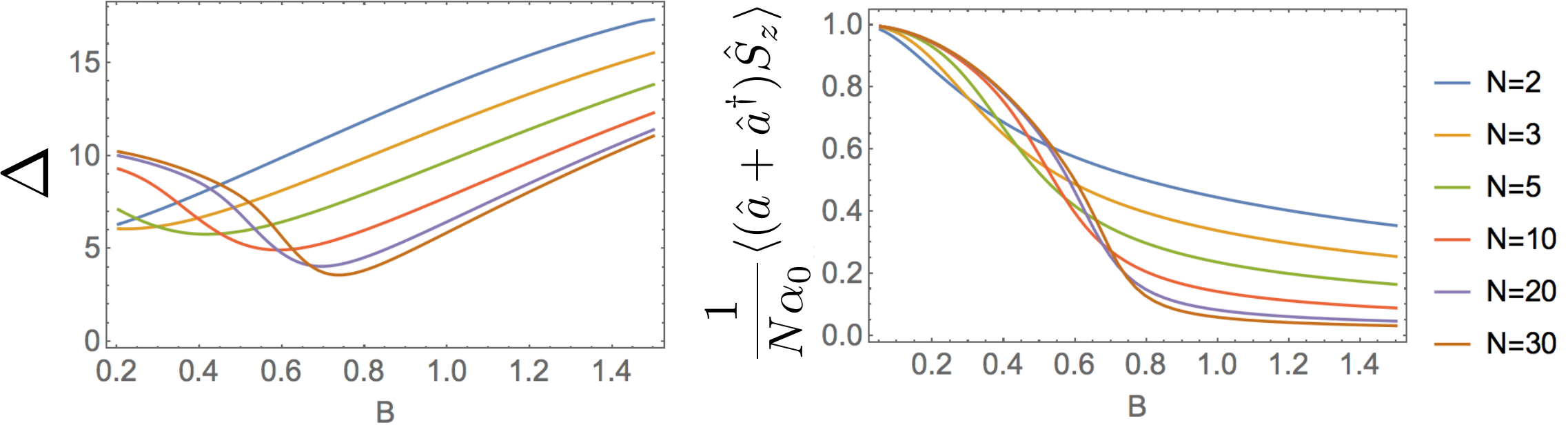}
 \includegraphics[width=16cm]{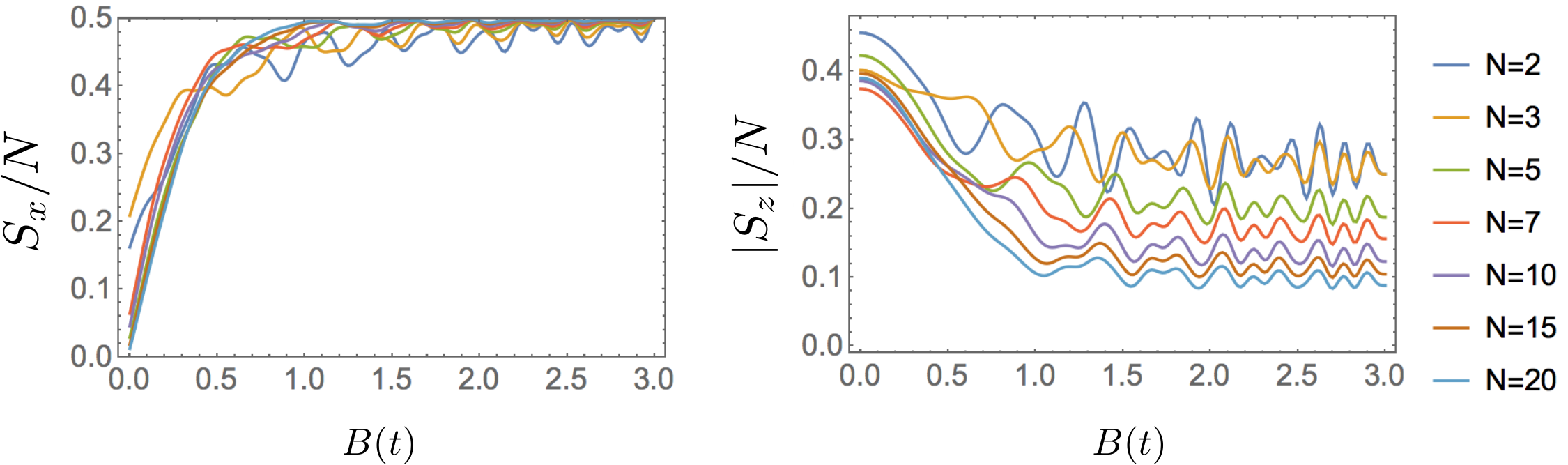}
 \caption{Key quantities and observables of Dicke model as a function of atom number $N$ and transverse field $B$. Energy gap $\Delta$ is between the ground-state and 
 excited state in the same parity sector. Magnetization $\vert S_z \vert$ and mean spin projection $S_x$ are computed for a LIN ramp with $t=2$ms and other parameters taken as per Fig.~2b of the manuscript.
 \label{fig:NPlot}} 
\end{figure*}

In this spirit, we plot the order parameter $\langle ( \hat{a} + \hat{a}^{\dagger} )\hat{S}_z \rangle$ and energy gap $\Delta$ between the ground-state and excited state in the same parity 
sector, for various ion numbers in Fig.~\ref{fig:NPlot} and as a function of transverse field strength $B$. A minimum in the energy gap, as a function of $B$, emerges for $N\gtrsim5$. This minimum 
is associated with the crossover between the normal and superradiant phase, and thus we predict that features of the crossover should be observable for $N\gtrsim 5$. 
This is consistent with the increasingly sharp transition observable in the order parameter for $N\gtrsim5$. Similarly, calculation of the spin observables $\vert S_z \vert$ and $S_x$ from dynamical 
ramps [plotted as a function of $B(t)$, parameters taken as per Fig.~2b of the manuscript], indicate that the crossover between normal and superradiant phases is evident for $N\gtrsim5$, 
which is easily satisfied by the experimentally considered crystal of $N\sim70$.

\section{Effect of the resonance on the energy gap \label{app:gap}}

\begin{figure}[!hbt]
    \includegraphics[width=8cm]{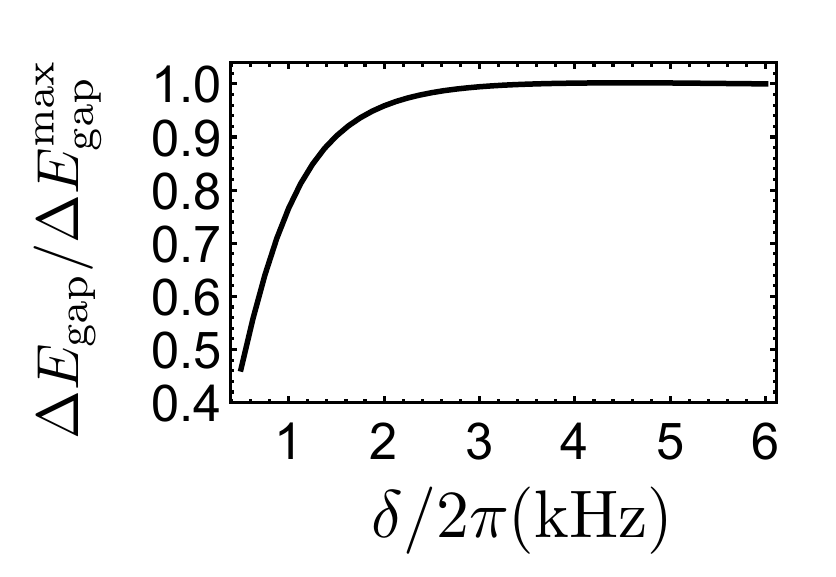}
    \caption{The size of the gap as a function of the detuning from the COM for $N=40$. As the size of the detuning increases, the resonant region of the Dicke model moves away from the
    quantum critical point separating the normal and superradiant phases. The energy gap at the critical point eventually saturates to a maximum value $\Delta E_{\rm gap}^{\rm max}$.  }
    \label{fig:norm_gap}
\end{figure}

As discussed in the main text, the Dicke Hamiltonian features a spin-boson resonance at $B=\vert \delta \vert $. At this field strength, the states $\vert m \rangle\vert -N/2 \rangle_x $ and
$\vert m-k\rangle\vert -N/2 + k \rangle_x$, with $k$ a positive integer, become nearly degenerate and can be resonantly coupled. The location of this resonance, relative to the critical field strength $B_c$,
can greatly affect the energy spectrum of the Dicke model and in particular the magnitude of the energy gap $\Delta$ between the ground-state and excited states in the same parity sector.
In this context, we can separate the effects of the resonance into two cases, defined by the relative position of the resonance to the critical field strength:

\begin{itemize}
\item {Case (i): $|\delta| \gg B_c$.} In this regime the resonance $B\simeq |\delta|$ is well separated from the critical point. The ground-state
$\vert \psi^{\rm Nor}_{0, N/2} \rangle = \vert 0 \rangle\vert -N/2 \rangle_x$ is decoupled from other states at resonance. Thus, the dynamics can be affected by resonant couplings to other states
(as above) only if excited states have become occupied throughout the quench.

\item{Case (ii): $|\delta| \sim B_c$.} If the resonance is in the proximity of the quantum critical point then the low-lying excitations near the critical point of the Dicke
Hamiltonian are non-trivial superpositions of spin and phonon excitations.  A radical consequence of this complex interplay is the relative reduction of the energy gap between the ground and
the first excited states of the same parity at the critical point. We illustrate this in Fig.~\ref{fig:norm_gap} as a function of detuning $\delta$, with the spin-phonon coupling $g_0$ scaled
such that the critical field strength $B_c = g_0^2/\delta$ is held fixed.
\end{itemize}

\section{Additional sequence to disentangle the spin cat-state \label{app:cat}}
In the main text, we briefly outline a procedure to disentangle the pure spin-cat state from adiabatic preparation of the ground-state of the Dicke Hamiltonian.
Here, we expand upon this discussion and give the appropriate details to verify this step.

In the weak-field limit, $B \ll B_c$, the ground-state of the Dicke Hamiltonian is the spin-phonon cat-state:
\begin{equation}
\vert \psi^{S}_{0,N/2} \rangle = \frac{1}{\sqrt{2}}\Big(\vert\alpha_0,0\rangle\vert N/2 \rangle_z \pm \vert-\alpha_0,0\rangle\vert -N/2 \rangle_z\Big) , \label{eqn:Supp_SpinPhCat}
\end{equation}
where $\alpha_0 = g_0\sqrt{N}/(2\delta)$. Without loss of generality we fix the sign of the superposition due to conservation of the spin-phonon parity symmetry, which dictates that the
positive superposition is prepared by an adiabatic quench from the strong-field ground-state $\vert \psi^{\mathrm{Nor}}_{0,N/2} \rangle$.

The choice of the sign in the superposition state Eq.~(\ref{eqn:Supp_SpinPhCat}) is dictated by the spin-phonon parity symmetry of the Dicke Hamiltonian. Specifically, $\hat{H}$ is
preserved under the simultaneous transformation of $\hat S_z\to -\hat S_z$, $\hat{S}_y \to -\hat{S}_y$ and $\hat a \to -\hat a$, and the associated conserved quantity of the Hamiltonian
is the generator of the symmetry $\hat{\Pi} \equiv e^{i\pi(\hat{a}^{\dagger}\hat{a} + \hat{S}_x + \frac{N}{2})}$. This symmetry dictates that when ramping
from high to low field, the state $\vert \psi_{0, N/2}^{Nor} \rangle$ will adiabatically connect to the superposition $\vert \psi_{0,N/2}^{S} \rangle$,
to conserve the parity $\langle \hat{\Pi} \rangle = e^{i\pi N}$. Specifically, for even $N$ the ground-state will be the
symmetric superposition with $\langle \hat{\Pi} \rangle = 1$, whilst for odd $N$ the ground-state is the anti-symmetric superposition
with $\langle \hat{\Pi} \rangle = -1$. Without loss of generality, we assume for the following that $N$ is even and thus we fix the sign of the
superposition to be positive.

Since the spin and phonon degrees of freedom are entangled in the ground-state [Eq.~(\ref{eqn:Supp_SpinPhCat})], the state obtained by tracing over the phonon degree of freedom is characterized by the reduced density operator
\begin{multline}
 \hat{\rho}_s = \frac{1}{2}\Big[ |N/2\rangle_z\langle N/2|_z + |-N/2\rangle_z\langle -N/2|_z \Big] \notag \\
  + \frac{e^{-|\alpha_0|^2}}{2} \Big[|-N/2\rangle_z\langle N/2|_z + |N/2\rangle_z\langle -N/2|_z \Big] .
\end{multline}
As the displacement amplitude $|\alpha_0|$ is increased, the reduced density matrix exponentially loses any information about the coherences which are exhibited in the spin-phonon superposition state.
As a concrete example, the ground-states of the main text typically have a mean phonon occupation $|\alpha_0|^2 \sim 2$--$30$ depending on the chosen parameters (i.e., detuning and spin-phonon coupling),
leading to $e^{-|\alpha_0|^2}\lesssim 0.1$. To fully probe the available coherences via only the spin degree of freedom, we must first transform Eq.~(\ref{eqn:Supp_SpinPhCat}) to a spin and phonon product state,
\begin{equation}
\vert \psi_{{\rm SB}} \rangle = \vert \phi\rangle \otimes \frac{1}{\sqrt{2}} \Big(\vert N/2 \rangle_z + \vert -N/2 \rangle_z\Big) , \label{eqn:Supp_SpinPhProduct}
\end{equation}
where $|\phi\rangle$ is some arbitrary state characterizing the phonon degree of freedom.

A possible procedure to achieve this decomposition is the following: At the conclusion of the ramp protocol, we fix the transverse field at $B=0$ and quench the detuning $\delta \rightarrow \delta^{\prime} = 2\delta$. The
spin-phonon state is then allowed to evolve for a duration $t_{d} = \pi/\delta^{\prime}$.
In the interaction picture, the initial spin-phonon superposition state evolves as
\begin{eqnarray}
 \vert \psi_{{\rm SB}} \rangle =  \hat{U}(t) \vert \psi_{0,N/2}^{S} \rangle ,
\end{eqnarray}
where
\begin{eqnarray}
  \hat{U}(t) & = & \hat{U}_{SB}(t)\hat{U}_{SS}(t) , \\
  \hat{U}_{\rm SS}(t) & = & \exp\left( -i \frac{J}{N} \hat S_z^2 t\right) , \\
 \hat{U}_{\rm SB}(t)& = & \hat D( \beta(t,\delta') S_z) .
 \end{eqnarray}
Here, $\hat U(t)$ is the propagator corresponding to the Dicke Hamiltonian with $B=0$ [Eq.~1 of the main text].
The propagator is comprised of two parts, the spin-spin propagator $\hat U_{\rm SS}(t)$ and the spin-phonon propagator $\hat U_{\rm SP}(t)$ where $\beta(t,\delta)= -g_0(1-e^{-i \delta t})/(2\delta\sqrt{N})$ (see \cite{Wall2017}
for a more detailed discussion).

\begin{figure}[!]
    \includegraphics[width=8cm]{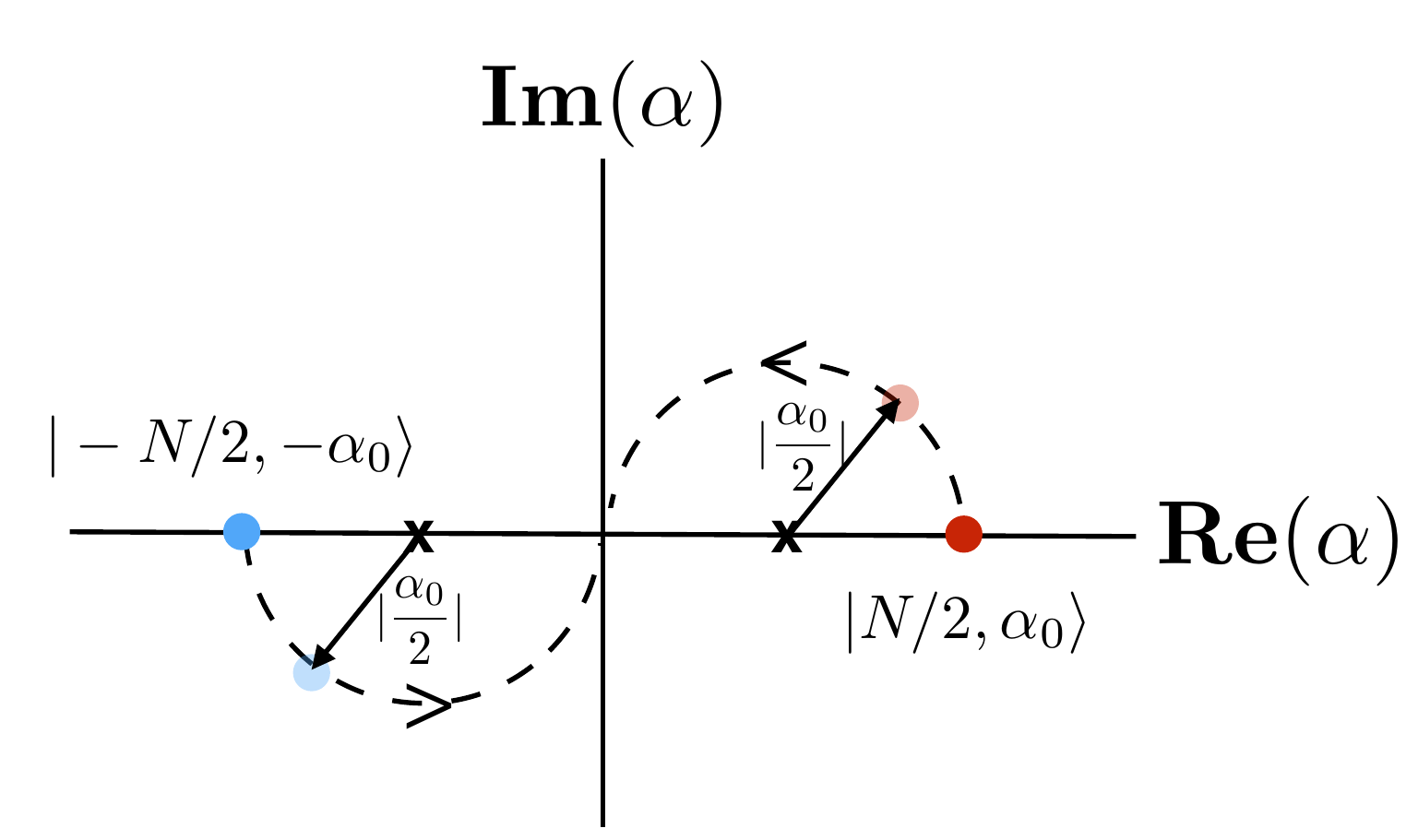}
    \caption{Schematic of the disentangling protocol to extract a pure spin cat-state from the spin-phonon
 ground-state $\vert \psi_{0,N/2}^{S} \rangle$. At the end of the ramp, we quench the detuning $\delta \rightarrow 2\delta$ and evolve the system for an additional duration $t_d = \pi/|2\delta|$ at fixed $B=0$.
 The phonon states start at opposing coherent amplitudes and undergo a spin-dependent coherent displacement which maps them to the phonon vacuum state.}
    \label{fig:Supp_disentangle}
\end{figure}

If at the end of the ramp we quench the detuning to $\delta^{\prime} = 2\delta$ and apply $\hat U(t)$ for $t_d=\pi/\delta^{\prime}$, such that $\beta(t_d,\delta')=-g N/(2\delta)$, it is then clear that $\hat{U}_{SB}$ will
displace the phonon coherent states (in a direction dependent on the sign of the $S_z$ component) back to vacuum, $|\pm\alpha_0,0\rangle \rightarrow |0\rangle$. We illustrate this displacement in Fig.~\ref{fig:Supp_disentangle}.
Note that the action of $\hat{U}_{SS}$ on the spin component of
the ground-state imprints an irrelevant global phase $\varphi =  JNt_d/2$ on the decoupled state Eq.~(\ref{eqn:Supp_SpinPhProduct}).

An alternative, but closely related, procedure to disentangle the spin-phonon state is to drive the spin-phonon coupling on resonance, $\delta \to \delta^{\prime} = 0$. In this case,
one must shift the phase of the drive by $\pi/2$ such that the spin-phonon coupling transforms as $\frac{g_0}{\sqrt{N}}(\hat{a} + \hat{a}^{\dagger})\hat{S}_z \to \frac{ig_0}{\sqrt{N}}(\hat{a} - \hat{a}^{\dagger})\hat{S}_z$, and subsequently evolve the system for a duration
$t_d = 1/|\delta|$. Following this procedure results in a spin-dependent coherent displacement of the phonon state back to vacuum, $|\pm\alpha_0,0\rangle \rightarrow |0\rangle$, in a manner similar to the previously discussed protocol.

We make one further point regarding the disentangling protocols. In the experimental system we generally characterize the initial state of the phonons as a thermal ensemble $\hat{\rho}_{\bar n}$ while the spin-degree of freedom is
prepared in a pure state, such that the initial spin-phonon state is $\hat{\rho}_{SB}(0) = \hat{\rho}_{\bar n} \otimes \vert -N/2 \rangle_x \langle -N/2 \vert_x $. If the protocol is adiabatic and there is no coupling between the
excited energy levels, then not only is the ground-state component of this initial ensemble mapped to the weak-field ground-state of the Dicke Hamiltonian, but the excited fraction due to the thermal distribution is also mapped identically.
This implies that the final state at the end of the ramp protocol will be a mixture of the true ground-state and the low-lying excitations, which, if $\delta^2 < g^2 N$, can be characterised as
displaced Fock states $\vert \pm\alpha_0, n \rangle$ where $n$ corresponds to the number of phonon excitations above the true ground-state.

The action of this protocol on these states is to identically displace the phonon state such that $\vert \pm\alpha_0, n \rangle \rightarrow |n\rangle$.
This maps the spin-phonon excited states to the form of a product state identical to Eq.~(\ref{eqn:Supp_SpinPhProduct}). Hence, tracing the phonons out of these excited
states also recovers the spin cat-state.

\section{Qualitative effects of initial phonon occupation}
\begin{figure}[!]
 \includegraphics[width=8cm]{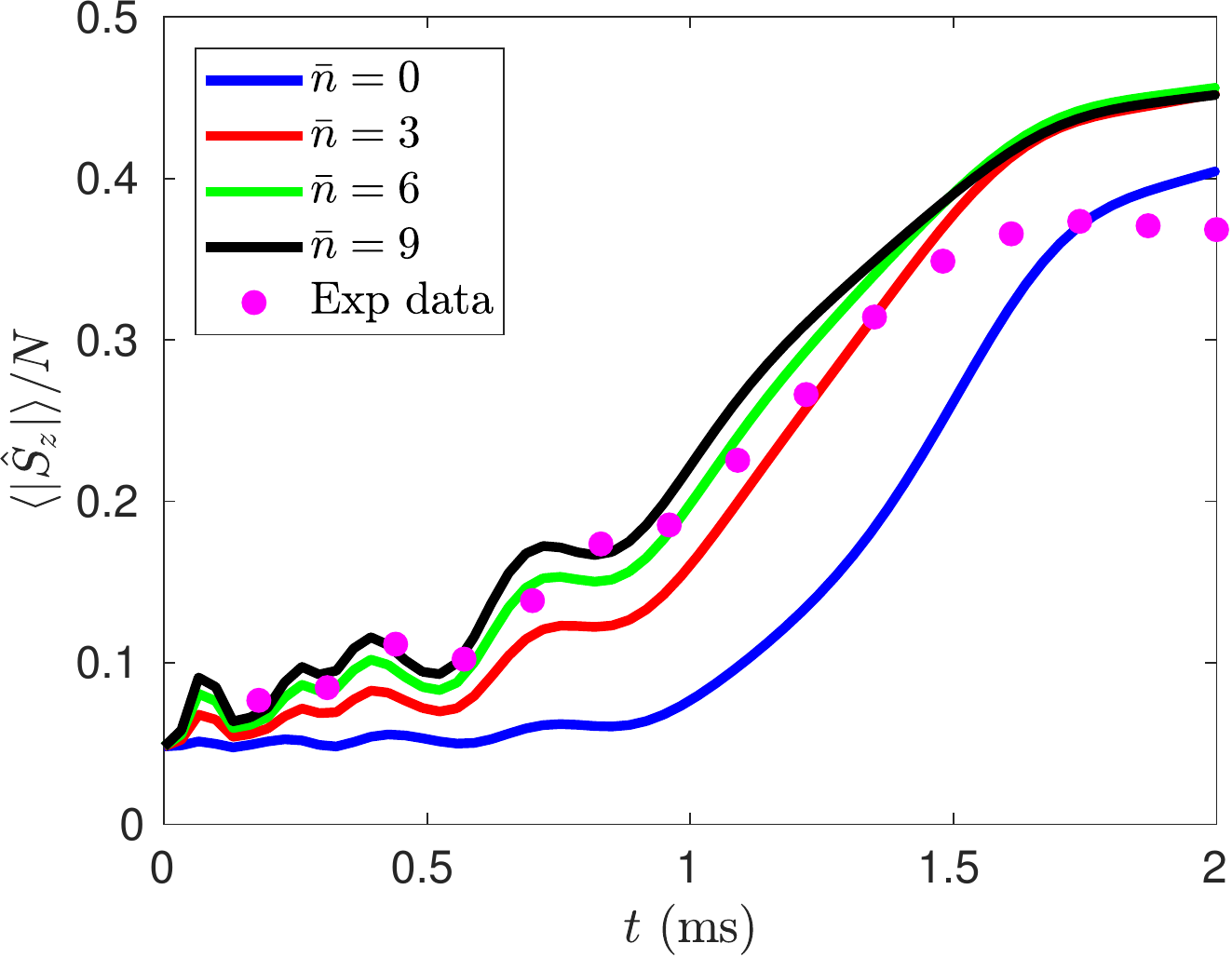}
 \caption{Comparison of magnetization $\langle |\hat{S}_z|\rangle$ from experimental data and theoretical calculations for different initial thermal occupation $\langle \hat{a}^{\dagger}\hat{a}\rangle = \bar{n}$ of the phonon mode. 
 The amplitude of the  oscillations at $t \lesssim 1$ clearly increase with $\bar{n}$, whilst the frequency appears to remain comparitively fixed. Data is for an EXP ramp, with all other parameters taken as per Fig.~2b of the 
 manuscript. \label{fig:PhononOsc}}
\end{figure}

In the main text we comment that the oscillations in the spin observable $\langle \vert \hat{S}_z \vert \rangle$ at short times is an indication of a non-negligible initial thermal occupation of the phonon mode (Fig.~2 of main text). 
Here, we support this conclusion by comparing results of theoretical calculations with different initial phonon occupation. Taking relevant parameters as per Fig.~2 of the main text and considering only the EXP ramp for simplicity, 
we plot the theoretical results for evolution of $\langle \vert \hat{S}_z \vert \rangle$ in Fig.~\ref{fig:PhononOsc}. We observe that if the phonons are taken to be initially in a vacuum state, the short time dynamics displays only extremely weak signs of 
oscillations. In contrast, when the phonons are taken to be initially described by a thermal ensemble with mean occupation $\bar{n} = 3$-$9$ there are signficant oscillations at short-times, consistent with the observed experimental data. Moreover, 
the final magnetization at the conclusion of the ramp protocol is much larger than that predicted from the vacuum case. The various values of $\bar{n}$ plotted give relatively similar agreement with the experimental data. However, 
$\bar{n} = 6$ is chosen in the main text as this is consistent with the estimated limit from Doppler cooling in the experiment.

\section{Inference of spin-phonon correlations \label{app:spinphonon}}
As detailed in the main text, we infer the presence of spin-phonon correlations from the time evolution of the spin observable $\langle \hat{S}_x \rangle$. Specifically, starting from
the Lindblad master equation for the density matrix of the spin-phonon system $\hat{\rho}$,
\begin{equation}
 \frac{d\hat{\rho}}{dt} = -\frac{i}{\hbar} \left[ \hat{H}^{\mathrm{Dicke}}, \hat{\rho} \right] +\frac{\Gamma_{el}}{2}\sum_{i=1}^N \left( \hat{\sigma}^z_i \hat{\rho} \hat{\sigma}^z_i - \hat{\rho} \right) ,
\end{equation}
wherein we have assumed single-particle dephasing is the dominant decoherence mechanism, it then follows that
\begin{equation}
 \frac{d\langle \hat{S}_x \rangle}{dt} = \frac{g_0}{\sqrt{N}}\langle \left( \hat{a} + \hat{a}^{\dagger} \right) \hat{S}_y \rangle - \Gamma_{el}\langle \hat{S}_x \rangle .
\end{equation}
From here it is straightforward to rearrange for the relation between the spin-phonon correlation and the evolution of $\langle \hat{S}_x \rangle$:
\begin{equation}
 \mathcal{C}_{\mathrm{sp-ph}} \equiv \langle \left( \hat{a} + \hat{a}^{\dagger} \right) \hat{S}_y \rangle = \frac{\sqrt{N}}{g_0} \left( \Gamma_{el}\langle \hat{S}_x \rangle + \frac{d\langle \hat{S}_x \rangle}{dt} \right) . \label{eqn:SpinPhCorr}
\end{equation}

We emphasize that evaluation of this spin-phonon correlation directly from either ground-state $\vert \psi^{\mathrm{Nor}}_{0,N/2} \rangle$ $\vert \psi^S_{0,N/2} \rangle$ yields $\mathcal{C}_{\mathrm{sp-ph}} = 0$, and this
result has been confirmed numerically for all transverse field strengths $B$ for the systems considered in the main text. This directly implies that the finite value reported in the main text is due to contributions
from excited states. Such contributions may come from diabatic excitations created throughout the ramping protocol or from the initial thermal phonon ensemble.

In the main text, we extract the spin-phonon correlation from the experimental data using the RHS of Eq.~(\ref{eqn:SpinPhCorr}) and evaluating the time-derivative numerically with a one-sided derivative. We
model dephasing using $\langle \hat{S}_x \rangle_{\Gamma} = \langle \hat{S}_x \rangle_{\Gamma = 0} e^{-\Gamma t}$ in our theoretical calculations, and extract the theoretically predicted spin-phonon correlation in an
identical manner.

\section{Experimental Optimisation of ramp protocols \label{app:RampOpt}}
To experimentally optimize the ramp protocols demonstrated in this work, we chose to optimize with respect to the total magnetization $\langle |\hat{S}_z| \rangle$ at the end of the ramp.
For the EXP ramp, we compared approximately $20$ different ramp profiles that utilized different exponential decay rates. Specifically, we would perform an experiment where the effective
transverse field was ramped from the initial field $B(t=0)$ at a fixed decay rate to $B\approx0$, where we then measured the spin-projection $M_z^{\mathrm{exp}}$ along the $\hat{z}$-axis.
This experiment was repeated, typically $500-700$ times, to gather statistics on the resulting distribution and obtain a measurement of $\langle|\hat{S}_z|\rangle$ from the histogram of $M_z^{\mathrm{exp}}$ measurements.
We then picked a ramp profile with a different exponential decay rate, and repeated this procedure. After identifying the exponential decay rate that optimized the final magnetization $\langle|\hat{S}_z|\rangle$,
we performed experiments that measured the magnetization distribution $P(M_z^{\mathrm{exp}})$ when stopping the ramp at different times, as discussed in the main text.

\begin{figure}[!]
    \includegraphics[width=8cm]{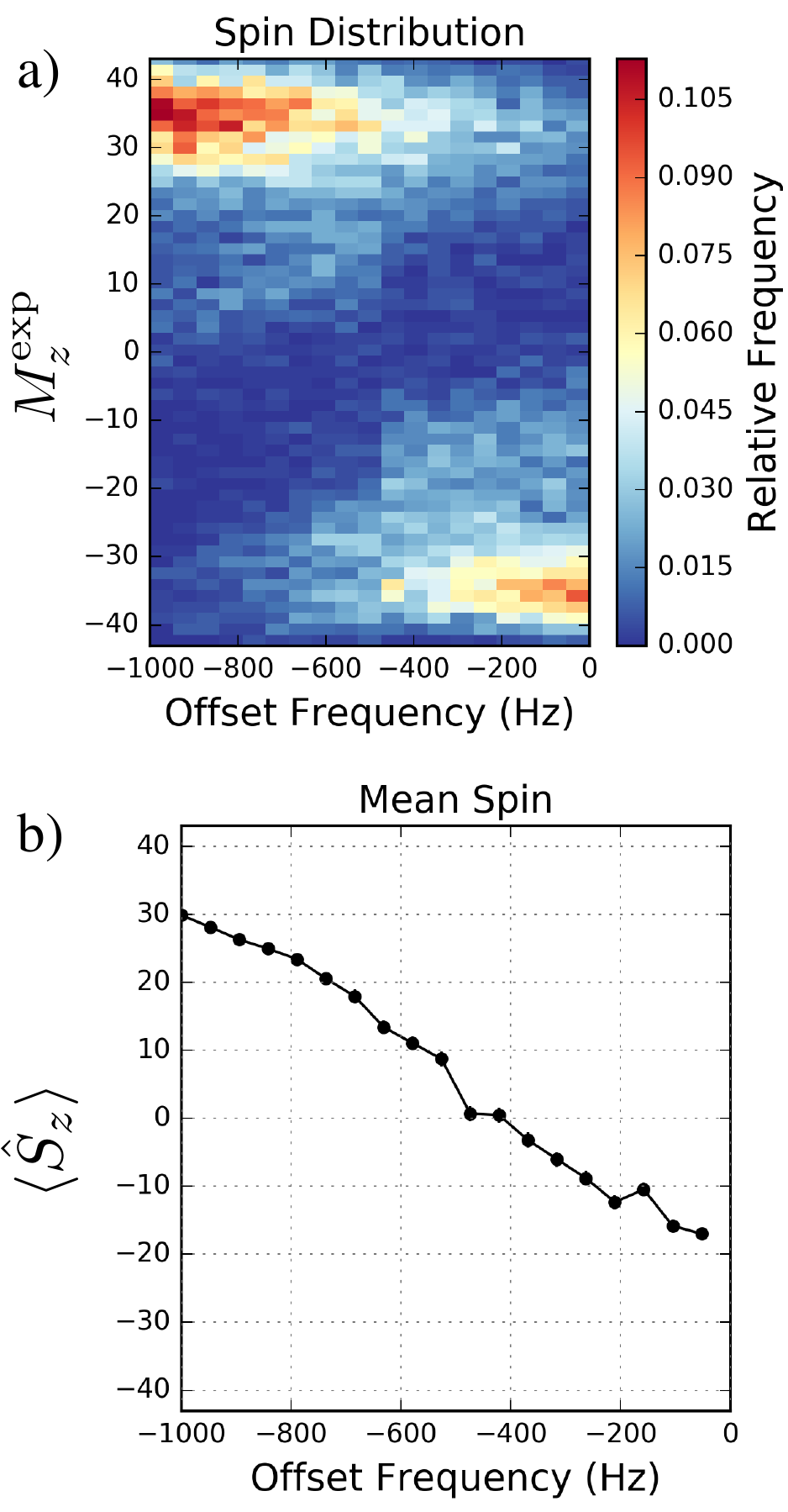}
    \caption{Balancing the $P(M_z^{\mathrm{exp}})$ distributions. (a) $P(M_z^{\mathrm{exp}})$ distribution functions extracted from experimental measurements of the spin-projection $M_z^{\mathrm{exp}}$ at the end of an EXP ramp of the transverse magnetic field to zero. The distribution functions are plotted as a function of frequency offset of the microwaves that generate the effective transverse magnetic field from the spin-flip resonance in the absence of the spin-dependent force. (b) Plot of the average magnetization $\langle\hat{S}_z\rangle$ from (a) as a function of the microwave offset frequency. An offset frequency that balanced the distributions at the end of the ramping sequence, defined by $\langle\hat{S}_z\rangle$, was used in studies described in the main text that measured the spin-projection distribution when stopping the ramp at different times.}
    \label{fig:Supp_Exp}
\end{figure}

When performing these ramp sequences and observing the distributions of $M_z^{\mathrm{exp}}$, in some cases the distributions would be biased to positive or negative spin-projection.
This can be observed in the distribution of Fig.~\ref{fig:Supp_Exp}(a) at zero offset frequency. Such an effect can be explained by a small longitudinal magnetic field that breaks the symmetry of the ground state.
The small longitudinal field was likely due to imperfect nulling of the Stark shift from the off-resonant laser beams that
generate the spin-dependent force \cite{Bohnet2016}. We would observe that this effect varies day to day. To compensate for this effect, during the ramp we would apply a small frequency offset to the microwaves that
provided the effective transverse field. For each frequency offset, we would measure the distribution of measurements $M_z^{\mathrm{exp}}$ at the end of the transverse field ramp as shown in Fig.~\ref{fig:Supp_Exp}(a).
For the appropriate offset, the distribution would be balanced, with large, separated peaks at positive and negative values of $M_z^{\mathrm{exp}}$.  To choose the optimum, we plot $\langle\hat{S}_z\rangle$ as a function of the
frequency offset and extract the zero crossing, as shown in Fig.~\ref{fig:Supp_Exp}(b).

\end{document}